\documentclass[11pt]{article} 
\usepackage{latexsym}
\usepackage{psfig}
\def\build#1_#2^#3{\mathrel{
\mathop{\kern0pt#1}\limits_{#2}^{#3}}}
\def\la{\mathrel{\mathpalette\fun <}}

\def\fun#1#2{\lower3.6pt\vbox{\baselineskip0pt\lineskip.9pt
        \ialign{$\mathsurround=0pt#1\hfill##\hfil$\crcr#2\crcr\sim\crcr}}}
\def\ben{\begin{equation}}
\def\be#1{\begin{equation}\label{eq:#1}}
\def\ee{\end{equation}}
\def\EC#1{(\ref{eq:#1})}

\def\ldb{\lambda_{\rm deB}}
\def\bx{{\bf x}}
\def\bv{{\bf v}}
\def\fxp{{\cal F}(\bx,\bv)}
\def\sch{Schr\"odinger~}
\def\beg{\begin{equation}}
\def\dist{distribution~}
\def\PS{Plummer sphere~}
\begin{document}

\title{Test-Bed Simulations of Collisionless, Self-Gravitating Systems
Using the Schr\"odinger Method}
\author{George Davies and Lawrence M. Widrow \\
Department of Physics, Queen's University, Kingston, K7L 3N6, Canada}

\maketitle
  
\begin{abstract}

The Schr\"odinger Method is a novel approach for modeling numerically
self-gravitating, collisionless systems that may have certain
advantages over N-body and phase space methods.  In particular,
smoothing is part of the dynamics and not just the force
calculation.  This paper describes test-bed simulations which
illustrate the viability of the Schr\"odinger Method.  We develop the
techniques necessary to handle ``hot'' systems as well as spherically
symmetric systems.  We demonstrate that the method can maintain a
stable, equilibrium star cluster by constructing and then evolving a
Plummer sphere.  We also consider nonequilibrium initial conditions
and follow the system as it attempts to reach virial equilibrium
through phase mixing and violent relaxation.  Finally we make a few
remarks concerning the dynamics of axions and other bosonic dark
matter candidates.  The Schr\"odinger Method, in principle, provides
an exact treatment of these fields.  However such ``scalar field''
simulations are feasible and warrented only if the deBroglie
wavelength of the particle is comparable to the size of the system of
interest, a situation that is almost certainly not the case for axions
in the Galaxy.  Therefore the Schr\"odinger Method treats axions in
the same way as other collisionless particles.  We challenge recent
claims in the literature that axions in the Galaxy form soliton stars.

\end{abstract}

\section{Introduction}

Recently Widrow and Kaiser (1993) have developed a new numerical
method for modeling self-gravitating, collisionless systems.  Known as
the Schr\"odinger Method (SM) this approach allows one to follow the
evolution of such systems on a three-dimensional Eulerian grid and
provides an alternative to both N-body and phase space methods.

Collisionless systems are ones in which the constituent particles move
under the influence of the mean gravitational potential generated by
all of the other particles.  The state of a collisionless system is
specified by a distribution function $f=f(\bx,\bv,t)$ which gives the
density of particles in phase space as a function of time.  $f$ is
treated as a continuous fluid whose evolution is
governed by the coupled Vlasov and Poisson equations (see e.g., Binney
\& Tremaine, 1986):

\be{vlasov} 
\frac{\partial f}{\partial t}=\sum_{i=1}^{3}
\left ( \frac{\partial V}{\partial x_i}
\frac{\partial f}{\partial v_i}-v_i \frac{\partial f}{\partial
x_i}\right )
\ee 
\be{poisson} 
\nabla^2 V = 4\pi G\int d^3 v f ~.
\ee

\noindent The Vlasov-Poisson pair must be solved numerically for most
time-dependent systems of interest.  One approach is to evolve $f$
directly in phase space though efforts along these lines have had
limited success primarily due to the large number of phase space
dimensions typically involved.  In contrast N-body or Particle methods
(see, e.g., Hockney \& Eastwood 1988) have been used successfully in
virtually all astrophysical problems involving gravitational dynamics.
The number of particles employed in an N-body simulation is typically
many orders of magnitude less than the actual number of particles in
the physical system one is modeling.  This discrepancy can lead to
unphysical effects due to ``particle noise'' such as two-body
relaxation.  Smoothing techniques alleviate this problem but at the
cost of spatial resolution and care must be taken to choose a
technique appropriate to the problem one is solving (Earn
\& Sellwood, 1995).  On the other hand phase space methods, by
explicitly constructing the smooth distribution function, avoid these
difficulties.

The SM attempts to circumvent the problems of both phase space and
N-body methods by encoding phase space information in a continuous
position space function which we call $\psi$.  Given an initial
distribution function at time $t_i$ we can find the distribution
function at a later time $t_f$ by the following general procedure:

\be{mapping}
f(\bx,\bv,t_i)\buildrel {\cal M} \over \Longrightarrow
\psi(\bx,t_i)\buildrel \rm DEM \over \Longrightarrow
\psi(\bx,t_f)\buildrel {\cal M}^* \over \Longrightarrow f(\bx,\bv,t_f)~.
\ee 

\noindent Here DEM is the dynamical equation of motion for $\psi$, ${\cal
M}$ maps $f$ to $\psi$, and ${\cal M}^*$ maps $\psi$ to $f$.  Ideally
${\cal M}$ would be an invertable map with ${\cal M}^*={\cal M}^{-1}$
in which case the procedure would yield an exact solution to the
Vlasov-Poisson pair.

With the SM, $\psi$ is formally identified as a complex
Schr\"odinger field obeying the coupled Schr\"odinger and Poisson
equations:

\be{schroed}
i\hbar\frac{\partial\psi}{\partial t}=
-\frac{\hbar^2}{2m}\nabla^2\psi + mV\psi
\ee

\be{poisson2}
\nabla^2 V=4\pi G\rho~.
\ee

\noindent ${\cal M}^*$ is the coherent state or Husimi transform (Husimi 1940), 
essentially the square of a windowed Fourier transform, and ${\cal M}$
is a procedure to sample the initial distribution function to be
discussed below.

In N-body simulations fictitious ``superparticles'' are used to
provide a statistical representation of the distribution function.
Imagine dividing position space into cells smaller than the scales of
interest but larger than the interparticle spacing.  The number of
particles in each cell gives the (course-grained) density field while
their velocity distribution provides the remaining information
contained in $f$.  The resolution in position space relative to
velocity space can be controlled by adjusting the cell size.  However
the size of a phase space resolution element, $\Delta\Omega$ is set by
the number of particles $N$ in the simulation:
$\Delta\Omega\simeq n\Omega/N$.  Here $\Omega$ is the volume in phase
space filled by the system (6 dimensional in general; 3 dimensional
for spherically symmetric systems with angular momentum (see below))
and $n$ is the number of particles required to have a reasonably
accurate estimate of $f$ at a given phase space point.

The SM shares some of these features.  The de Broglie wavelength
$\ldb$ for $\psi$ enters as a model parameter and is chosen to be as
small as possible, in general a few grid spacings.  Position space is
divided into regions smaller than the scales of interest but larger
than $\ldb$ so that the WKB approximation applies.  We can then think
of $\psi$ in this region as the superposition of plane waves each of
which represent different velocity streams in the distribution
function.  The correspondence principle guarantees that the
``acceleration'' of each stream is given by Newton's Law.  Since
$\psi$ is a continuous function there is no problem with particle
noise.  Moreover it is a function only of position
rather than the phase space coordinates,
and is therefore easier to work with. 
Formally, the resolution
in phase space is set by the uncertainty principle.  In practice it is
set by the number of grid points $N_g$: $\Delta\Omega\simeq n_g\Omega/N_g$ 
where $n_g$ is the number grid points across a typical de Broglie
wavelength.
We therefore expect the SM to be competitive with particle mesh 
simulations in terms of resolution in phase space as a function of the 
number of grid points.

In a previous letter Widrow and Kaiser (1993) outlined the SM and
presented simple illustrative simulations.  In this paper we expand on
that original work demonstrating, in a more quantitative way, the
validity of the method.  In particular we develop a new technique for
handling ``hot'' systems (e.g. virialized systems) as well as systems
with spherical symmetry.  In Section 2 we review the SM and describe
some of the new techniques implemented in this paper.  Section 3
presents results of test-bed simulations.  In particular we simulate a
stable equilibrium star cluster (Plummer sphere) and a nonequilibrium
cluster.  Both simulations assume spherical symmetry but include
angular momentum.  In Section 4 we make a few remarks concerning
recent work on axion dynamics and in particular challenge claims made
in the literature which suggest that the ``field'' nature of axions is
important for understanding their behaviour in the Galaxy.  Section 5 offers a
summary and some concluding remarks.

\section{Schr\"odinger Method}

\subsection{From $\psi$ to $\cal F$}

We use the coherent-state (Husimi 1940) representation to construct a
phase space distribution function $\fxp$ from $\psi$.  Mathematically
this is just the absolute square of a windowed Fourier
transform. However, the physical content of this mapping is clearer
when expressed in {\it bra-ket} notation:
 
\be{defF}
{\cal F}(\bx,\bv)=|\langle\eta(\bx,\bv)|\psi\rangle|^2\equiv
|\Psi(\bx,\bv )|^2~.
\ee

\noindent Here $|\eta(\bx,\bv)\rangle$ is the wavepacket for a single 
`particle' centered on the phase space point $(\bx,\bv)$,

\be{packet}
\langle\eta(\bx,\bv)|\bx'\rangle=
\left (\frac{1}{2\pi\hbar}\right )^{3/2}
\left (\frac{1}{\pi\eta^2}\right )^{3/4}
e^{-\left (\bx-\bx'\right )^2/2\eta^2-im\bv\cdot\bx'/\hbar}~.
\ee

\noindent Equation \EC{defF} can then be seen as a 
prescription for decomposing $\psi$ into
Gaussian wavepackets with velocity $\bv$ and position $\bx$.
\noindent One can show, directly from the Schr\"odinger equation, that
${\cal F}$ satisfies

\be{qvlasov} 
\frac{\partial {\cal F}}{\partial t}=\sum_{i=1}^{3}
\left ( \frac{\partial V}{\partial x_i}
\frac{\partial {\cal F}}{\partial v_i}-v_i 
\frac{\partial {\cal F}}{\partial x_i}\right )
+\left [O\left (\frac{\eta^2}{L^2}\right )+
O\left (\frac{\ldb^2}{\eta^2}\right )\right ]
\frac{\partial {\cal F}}{\partial t}
\ee 

\noindent (see, e.g., Skodje, Rohrs, \& van Buskirk 1989)
where $L$ is the length scale over which the system varies and
$\ldb\sim|\psi/\nabla\psi |\simeq \hbar/mv$ is the typically de
Broglie wavelength of $\psi$.  This equation reduces approximately to
the Vlasov equation provided $\ldb\ll\eta\ll L$.

Moments of the  distribution function are calculated in the usual way
(e.g., $\langle v^n\rangle = \int d^3v v^n {\cal F}\big /
\int d^3v {\cal F}$).  In particular
the density field $\rho$, which is the source term for the potential $V$,
is given by
\be{smoothrho}
\rho=\int d^3 v{\cal F} = \left(\frac{1}{\pi\eta^2}\right)^{3/2}
\int d^3 \bx' e^{-\left (\bx-\bx'\right )^2/\eta^2}|\psi(\bx')|^2  
\ee

\noindent Evidently $\rho$ is found by applying a Gaussian filter to 
$|\psi|^2$, in effect, removing the high frequency modes.
Of course what we are interested in is the potential $V\sim
\nabla^{-2}\rho$ and, since the inverse Laplacian also removes high
frequency modes, we can safely replace $\rho$ with $|\psi|^2$ in
equation \EC{poisson2}.  In practice we calculate $V$ from $\rho$ using FFT.
Particle mesh simulations also use FFT to calculate the force from $\rho$ and
so the CPU time for this part of the calculation will be the same for the 
two methods.

\subsection{From $f$ to $\psi$}

We begin by sampling the distribution function, compiling a list of
phase space points much as we would in an N-body simulation.  The
initial $\psi$ is then taken to be the (incoherent) superposition of
wavepackets centered on each of these points:
\be{ansatz}
|\psi\rangle = \frac{1}{\sqrt{N}}\sum_{j=1}^{N}|\eta(\bx_j,\bv_j)\rangle~.
\ee
The $|\eta(\bx_j,\bv_j)\rangle$ 
are given by equation \EC{packet}.  The
phase factor $\exp{(i\varphi_j)}$  
is included to insure that the wavepackets will add
incoherently.  
While the physical motivation for this prescription is clear (we are just
summing up `particles') a more mathematical motivation is the completeness of
$|\eta(\bx,\bv)\rangle$:
\beg
\int d^3 {\bf x} \int d^3 {\bf v} |\eta(\bx,\bv)\rangle\langle\eta(\bx,\bv)|
= I ~.
\ee
To see that this form provides a valid representation
of $f$ we compute ${\cal F}(\bx,\bv)={\cal M}^*\left ({\cal M}f\right
)$:

\begin{eqnarray}
\label{ansatz}
{\cal F}(\bx,\bv)&=&\frac{1}{N}
\sum_{j=1}^N\sum_{k=1}^N
\langle\eta|\eta_k\rangle
\langle\eta_j|\eta\rangle\\
&=&\frac{1}{N}\sum_{jk}
e^{-\left (\bx-\bx_j\right )^2/4\eta^2}
e^{-\left (\bx-\bx_k\right )^2/4\eta^2}
e^{-m^2\eta^2\left (\bv-\bv_j\right )^2/\hbar^2}
e^{-m^2\eta^2\left (\bv-\bv_k\right )^2/\hbar^2}
\Gamma_{jk}
\end{eqnarray}
where
\be{gammajk}
\Gamma_{jk}=
e^{-2im\bx\cdot\left (\bv_j-\bv_k\right )/\hbar}
e^{-2im\bv\cdot\left (\bx_j-\bx_k\right )/\hbar}
e^{2i\left ( m\left (\bx_j\cdot\bv_j-\bx_k\cdot\bv_k\right
)/\hbar+(\varphi_j-\varphi_k)/2\right )}~.
\ee
Splitting this into a summation over equal and unequal indices we find,
\be{sminit}
{\cal F}=\frac{1}{N}\sum_{j=1}^N
e^{-\left (\bx-\bx_j\right )^2/2\eta^2}
e^{-2m^2\eta^2\left (\bv-\bv_j\right )^2/\hbar^2}
+{\cal F}_1 ,
\ee
where ${\cal F}_1$ is a summation over the interference terms and will
tend to be small.
It is instructive to compare this result with the usual N-body
\dist function:
\be{nbinit}
f(\bx,\bv)=\frac{1}{N}\sum_{j=1}^N\delta(\bx-\bx_j)\delta(\bv-\bv_j) .
\ee
In the limit where the interference terms vanish
equation \EC{sminit} becomes equation \EC{nbinit} with the delta
functions smoothed into Gaussians.

The discussion above reveals a key distinction between N-body methods
and the SM:  In an N-body simulation, smoothing is implemented only 
in the force calculation.  The `particles' in the SM are themselves 
treated as extended objects which evolve as would a distribution 
of particles (see below).  Indeed, the particles/wavepackets in the
initial distribution mix with one another and do not keep their 
identity as do particles in an N-body simulation.  We therefore
have, in effect, a form of {\it dynamical} smoothing.

\subsection{Spherical Symmetry}

We now specialize to spherical symmetry where phase space has
three dimensions; radial coordinate $r$, radial velocity $v_r$, and
angular momentum $j=r\sqrt{v_\theta^2+v_\phi^2}$.  
The Vlasov and Poisson equations take the form:
\be{sphvlasov}
\frac{\partial f}{\partial t}  ~+~
v_r \frac{\partial f}{\partial r} ~ +~
\left ( \frac{j^2}{r^3}-\frac{\partial V}{\partial r}\right )
\frac{\partial f}{\partial v_r}~=~0
\ee 
\be{sphpoisson} 
\frac{\partial}{\partial r}r^2\frac{\partial V}
{\partial r}~ = ~4\pi^2 G
\int_{-\infty}^\infty dv_r \int_0^\infty dj^2 f ~~.
\ee
The derivation of equation \EC{sphvlasov} assumes implicitly that $j\ne 0$.
(The Jacobian of the transformation from $(\bx,\bv)$ to $(r,v_r,j)$ is
singular for $j=0$).  This in general does not present a
problem so long as $j$ is treated as a continuous variable.  However
for numerical work $j$ is discretized and $j=0$ must be treated
with care.  
Of course $j$ can be viewed as a label since
no differential operation with respect to $j$ is ever performed.
We can therefore write $f(r,v_r,j)=f_j(r,v_r)$ and treat the evolution
of the different $f_j$'s separately.

For radial orbits ($j=0$) we write
\be{jeq0}
f_0(r,v_r)~\equiv~\frac{F(r,v_r)}{r^2}\delta(v_\theta)\delta(v_\phi)~.
\ee
where $F$ satisfies the following equation:
\be{jeq0vlsv}
\frac{\partial F}{\partial t}+
v_r\frac{\partial F}{\partial r}-
\frac{\partial V}{\partial r}
\frac{\partial F}{\partial v_r}=0
\ee
The Poisson equation is now
\be{poisson3}
\frac{\partial}{\partial r}r^2
\frac{\partial V}{\partial r}~=~
4\pi G\int_{-\infty}^{\infty}dv_r\left (F+\pi\sum_{j^2} f_j\right )~.
\ee
where the sum over $j^2$ is the discretized version of the integral
over $j^2$ in equation \EC{sphpoisson}.

It is straightforward to extend the SM method to spherically symmetric
systems with angular momentum.  We construct a $\psi_j$ for each 
$f_j$ in the initial distribution function:
For $j=0$ we have
\be{sphcsr0}
\Psi_0(r,v_r)~=~
C\int_0^\infty dr' e^{-\left (r-r'\right )^2/2\eta^2}
e^{-imv_r r'/\hbar}\psi_0(r')
\ee
where $\psi_0$ obeys the equation
\be{sphschroed0}
i\hbar\frac{\partial\psi_0}{\partial t}~=~
-\frac{\hbar^2}{2m}\frac{\partial^2}{\partial r^2}\psi_0~ +~ 
mV\psi_0~~~~~~~;~~~~~~~\left.\frac{\partial\psi_0}{\partial r}\right|_{r=0}~=~0~.
\ee
For $j\ne 0$
\be{sphcsr}
\Psi_j(r,v_r)~=~C\int_0^\infty dr' e^{-\left (r-r'\right )^2/2\eta^2}
e^{-imv_r r'/\hbar}\phi_j(r')
\ee
\be{sphschroed}
i\hbar\frac{\partial\phi_j}{\partial t}=
-\frac{\hbar^2}{2m}\frac{\partial^2}{\partial r^2}\phi_j ~+~ 
m\left (V+\frac{j^2}{2r^2}\right )\phi_j~~~~~~~;~~~~~~~
\left.\psi_j\right|_{r=0}~=~0~.
\ee
The distribution function is then
\be{sphdf}
r^2|\Psi_0|^2\simeq F~~~~~;~~~~|\Psi_j|^2\simeq f_j
\ee
and the Poisson equation becomes
\be{poisson4}
\frac{1}{r^2}\frac{\partial}{\partial r}r^2
\frac{\partial}{\partial r} V ~=~ 4\pi G\rho
\ee
where
\be{rho}
\rho~=~|\psi_0|^2 ~+ ~\frac{\pi}{r^2}\sum_{j^2}|\phi_j|^2~.
\ee
Equation \EC{rho} emphasizes an important point: the $j=0$ phase space map
gives a density while the $j\ne 0$ map produces $dM(j^2)/dr$.

\subsection{Numerical Preliminaries}

It is convenient to write our equations in terms of the dimensionless
quantities ${\bf y}=\bx/L$, $\tau=t/T$, $\chi=\psi/\sqrt{\rho_0}$, and
$U=\frac{T^2}{L^2}V$ where $L$ and $\rho_0$ are the characteristic size
and density of the system of interest.  We then have

\be{dmslsch}
i\frac{\partial\chi}{\partial\tau}=-\frac{\alpha}{2}\nabla_y^2\chi
+\frac{1}{\alpha}U\chi
\ee
\be{dmslpoi}
\nabla_y^2 U = \frac{3\pi^2}{32}\beta^2\chi\chi^*
\ee
where $\alpha\equiv\hbar T/\mu L^2$ and $\beta^2=32G\rho_0 T^2/3\pi$.
We therefore have
three dimensionless parameters, $\alpha$, $\beta$, 
and $\eta/L$, at our disposal. $\beta$ determines the choice of timescale:
$T=\beta T_d$ where $T_d=\sqrt{3\pi/32G\rho_0}$ is the dynamical time.
As we now show $\alpha$ determines the size of the phase space region
accessible in the simulation while $\eta/L$ sets the relative
resolution in position and velocity space.  For simplicity consider
one coordinate or two phase space dimensions.  Suppose that our
system comfortably fits inside a region of size ${\cal L}$ in units of
$L$.  We discretize position space by choosing $y_j=j{\cal L}/N$ where
$N$ is the number of gridpoints.  Further suppose that the system,
when viewed in velocity space, comfortably fits inside a region of
size ${\cal V}$ in units of $L/T_d$.  The lattice spacing in velocity
space is connected to the lattice spacing in position space in the
usual way (see, e.g., Press, Flannery, Teukolsky, \& Vetterling 1986),
$i.e.$ the velocity is proportional to the wave number of the FFT.
For discrete points in
velocity space we have (in units of $L/T$)  
$u_k=2\pi k \alpha /{\cal L}$ which implies that we should set
$\alpha=\beta{\cal VL}/2\pi N$. Writing $\alpha=\gamma{\cal L}/N$
where $\gamma=O(1)$ we find
$\beta=2\pi\gamma/{\cal V}$. 
In practice we find that $\gamma \simeq 5$ works well. For the \PS simulations
below we take
${\cal V}\simeq 5$
and $\beta\simeq 2\pi$.

The SM has the desirable feature that many quantities of interest can be
calculated without having to actually construct the full phase space \dist
function. We have already seen an example of this when calculating $\rho$.
In general, if the function does not depend on velocity, the velocity integral
will collapse into a delta function.
\begin{eqnarray}
\langle Q\rangle&=&\int d^3\bx~d^3\bv~Q(\bx)~\fxp \\
&=&(2\pi\eta^2)^{-3/2}\int d^3\bx~d^3\bx'~e^{-\left(\bx-\bx'\right)^2/\eta^2}
~Q(\bx)~|\psi(\bx')|^2\\
&\simeq&\int d^3\bx ~Q(\bx)~|\psi(\bx)|^2
\end{eqnarray}
In the case of velocity moments
we can replace the $v^n$ term with
a corresponding derivative, perform a change of coordinates and integrate by
part to get,
\beg
\langle v_j^n\rangle=\left(-i\frac{\alpha}{\beta}\right)^n
\int d^3\bx~
\left.\left[\frac{\partial^n~}{\partial s^n}
\left(e^{-s^2/4\eta^2}\psi(\bx+s\hat{\bf n}_j/2)\psi^*(\bx-s\hat{\bf n}_j/2)
\right)\right]\right|_{s=0} ,
\ee
where $\hat{\bf n}_j$ is the unit vector in the $j$ direction.
For the important case of $\langle v^2 \rangle$ this expression reduces to,
\be{vsq}
\langle v^2\rangle\simeq-\left(\frac{\alpha}{\beta}\right)^2
\int d^3\bx \left\{\psi(\bx)\nabla^2\psi^*(\bx)
\right\},
\ee
where we have neglected a term proportional to $\psi\psi^*$, and have 
integrated
by parts.
This is just the usual quantum mechanical result 
$\langle v^2\rangle=\langle \hat{P}^2\rangle/m^2$.

Applying this to the case of spherical symmetry we then find,
\begin{eqnarray}
\langle Q\rangle&\simeq&\sum_{j^2}\int_{0}^{\infty}dr~Q(r)~|\psi_j (r)|^2 \\
\langle v_r^2\rangle&\simeq&-\sum_{j^2}\left(\frac{\hbar}{m}\right)^2
\int_{0}^{\infty}dr~\psi_j^*(r)
\frac{d^2\psi_j (r)}{dr^2} .
\end{eqnarray}
One can, of course, resort to the `brute force' calculation of more complicated
quantities,
\beg
\langle Q\rangle= \sum_{j^2} \int dr \int dv_r~Q(r,v_r,j^2)~
{\cal F}_{j}(r,v_r).
\ee
The draw back of this approach is that it requires an $O(N^4\ln(N))$
calculation for each $j$ plane.
These calculations can be very time consuming, and it is time well spent 
looking for a short cut of the kind just presented.

\section{Test-Bed Simulations}

\subsection{Test Particles}

\begin{figure}
\psfig{file=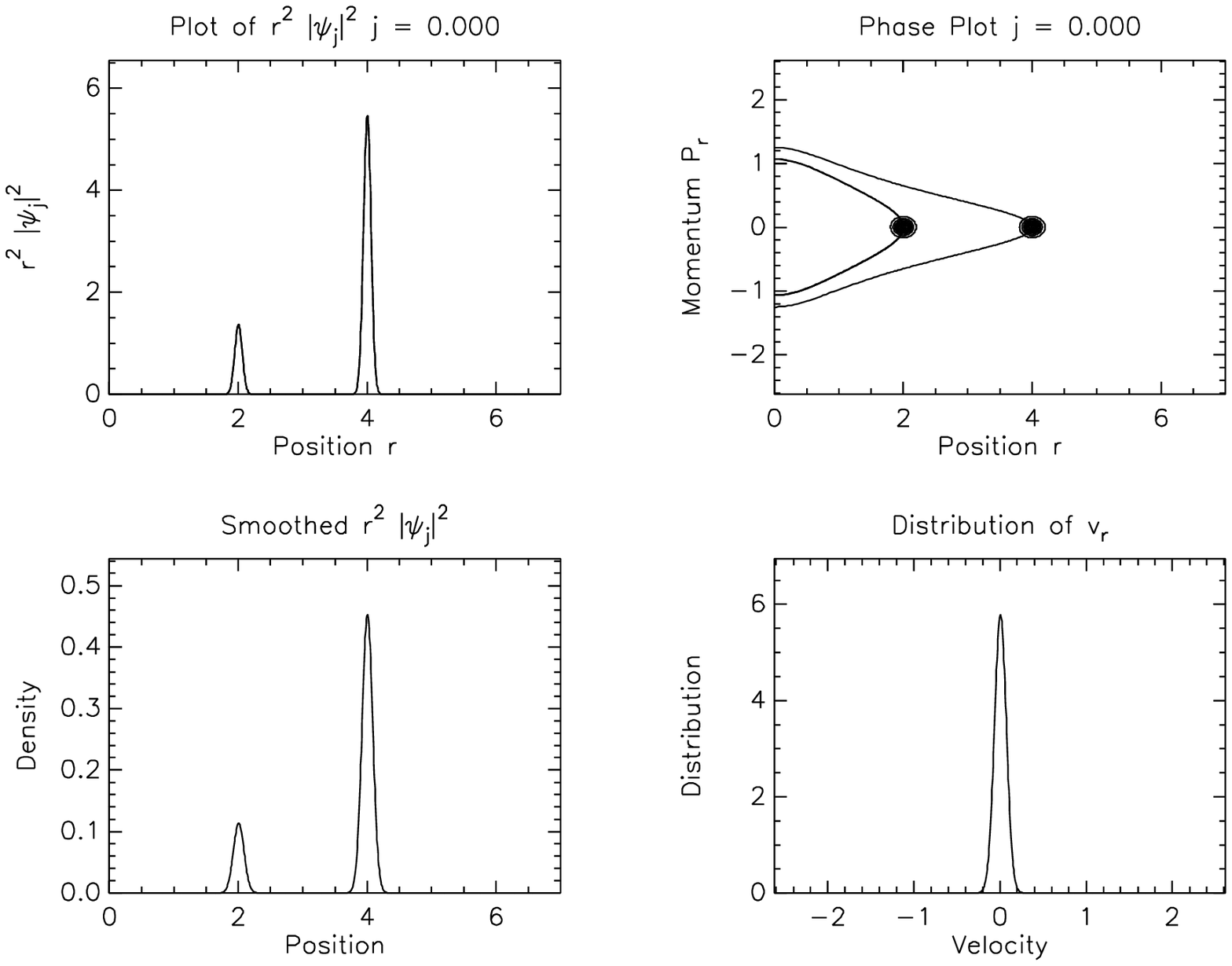,width=5.25in,angle=0}
\psfig{file=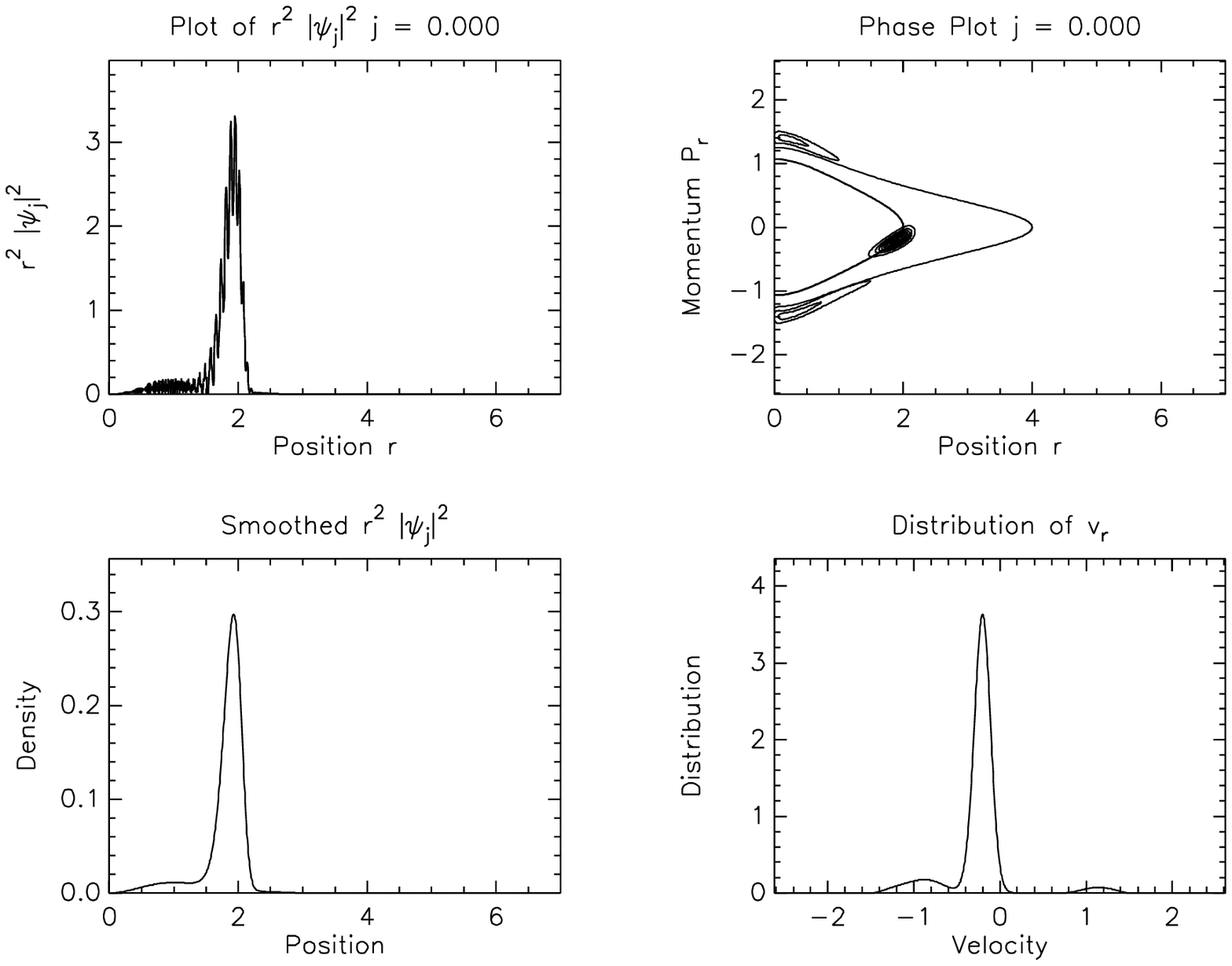,width=5.25in,angle=0}
\caption{}
\label{fig:particles1}
\end{figure}

\begin{figure}
\psfig{file=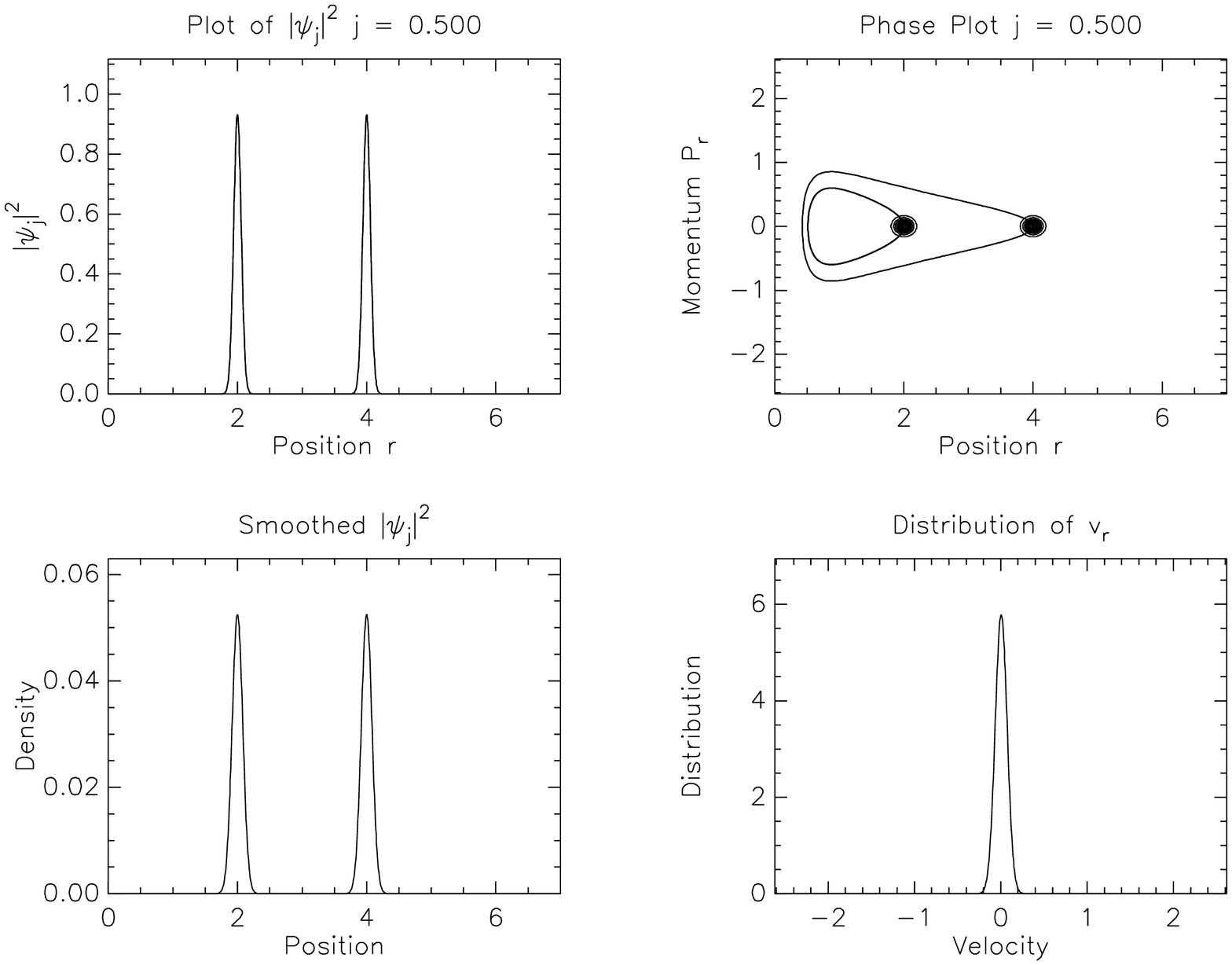,width=5.25in,angle=90}
\psfig{file=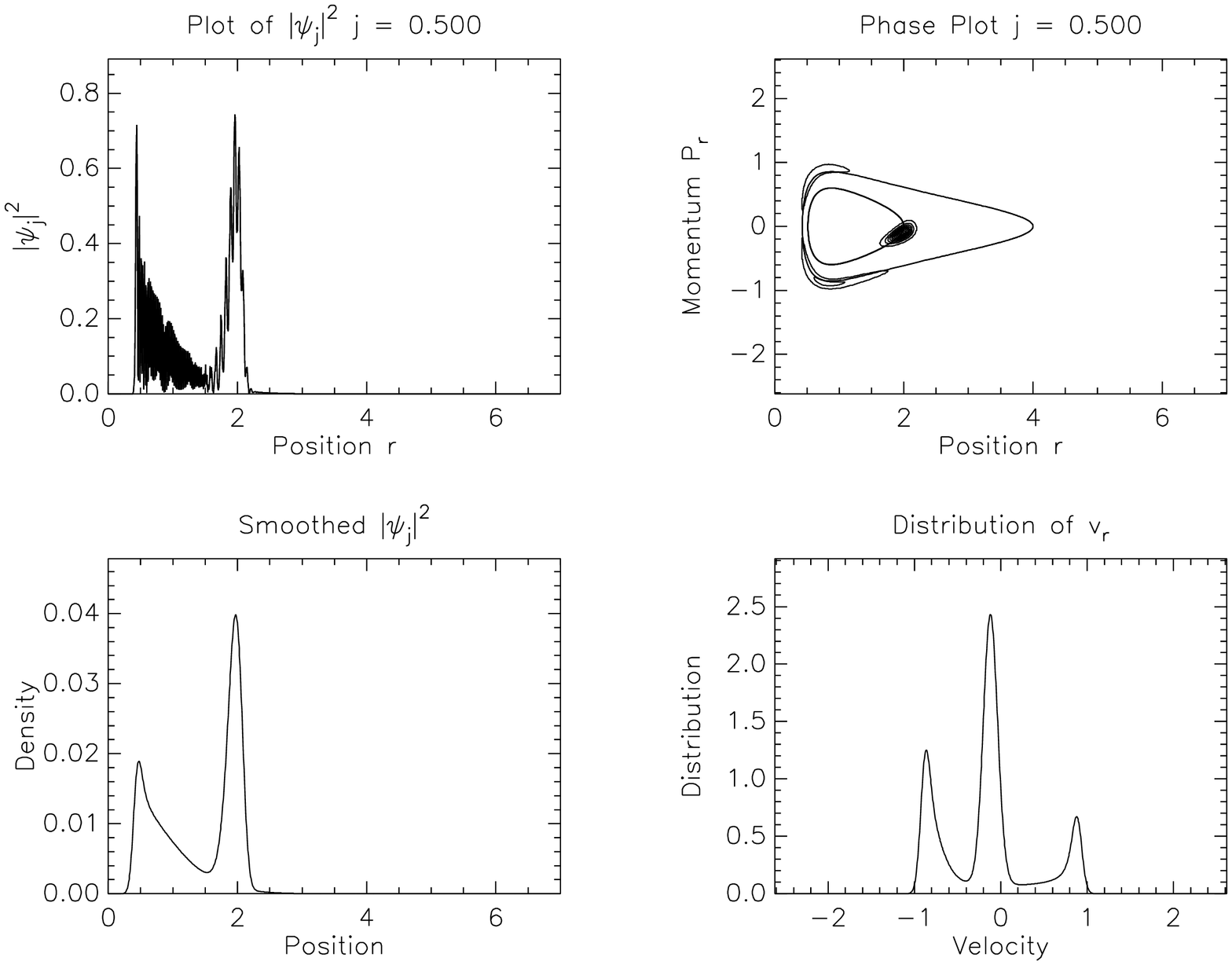,width=5.25in,angle=90}
\caption{}
\label{fig:particles2}
\end{figure}

As a first concrete example of the SM we will consider the motion of 
`test particles' in a fixed Plummer sphere potential,
\be{fixpot}
V(r)=-GM\left (L^2+r^2\right )^{-1/2} .
\ee
The SM analogue of a test particle
is a single coherent state wave packet centered at some chosen point in phase
space. Here we will examine the motion of four particles, two in the
$j=0$ plane and two in the $j=0.5$ plane. All four particles are taken to 
have zero initial radial velocity. The resulting initial wave functions
are then just the sum of two Gaussians centered at the particle positions.
The system is evolved by numerically solving
equations \EC{schroed} and \EC{fixpot}, using the algorithm
presented in Visscher (1991). This is an explicit algorithm and is at least
a factor of three faster than the usual implicit ones.
Figures \ref{fig:particles1} and \ref{fig:particles2}
show the four particles moving in the
Plummer sphere potential.
These plots help to show the physical 
connection between $\psi$ and ${\cal F}$, and the difference between the
$j=0$ and $j\ne 0$ phase space maps.
The particle positions are given by the peaks of the wave function while the
velocity information is encoded in the high frequency modes of $\psi$.
From these plots one can see that the particles do
follow the expected trajectories, justifying our choice of a spherical
phase space map.  

These plots also allow an interesting comparison of the SM and 
N-Body methods at a
fundamental level.  When one follows the evolution of an N-Body
`superparticle' it effectively carries with it a piece of phase space
that has a fixed size and shape. When modeling systems in which mixing
occurs these phase space blocks lead to unphysical coarseness of the
\dist function. The evolution of the \sch particles, on the other
hand, carry with them a piece of phase space that has a {\em variable}
shape and fixed volume.  The motion of the test particles in figure
\ref{fig:particles1} illustrates this point. The individual \sch
particles spread out along the phase space orbit in the same fashion
as an equivalent \dist of point particles would. The conclusion is
that fiducial SM particles can themselves take part in the phase
mixing process. 

We can also examine the evolution of an initially `cold' (i.e., single
velocity stream) distribution of test particles.
The corresponding wave function is given by 
square root of the desired density multiplied by an appropriate phase factor:
\beg
\psi(\bx)=\sqrt{\rho(\bx)}e^{-i\theta({\bf x})}
\ee
where ${\bf \nabla}\theta=m\bv'$.
This leads to the phase space \dist
\beg
{\cal F}(\bx,\bv)\sim
\rho(\bx)e^{-\eta^2\left (m\bv'-\hbar{\bf 
\nabla}\theta\right )/\hbar^2}.
\ee
Compare this expression with the actual distribution function for a
cold system:
\beg
f(\bx,\bv)=\rho(\bx)\delta(\bv-\bv'({\bf x})) ~:
\ee
The delta function has
been replaced with a Gaussian of width $\sim\hbar/\eta$.

Figure \ref{fig:phasemix1} shows the phase space evolution of initially
cold ($v_r=0$) test-particles with angular momentum $j=0.5$ (in units
of $L/T_d$) moving in the 
fixed potential of a Plummer sphere.
The initial $\psi$ is given by
\beg
\psi(r)=\sqrt{
\left.\left[\int_{-\infty}^{\infty}dv_r~f(r,v_r,j^2)\right]\right|_{j=0.5}},
\ee
where $f$ is the Plummer sphere \dist function (see below).  With no velocity
pressure to support itself the system quickly begins to phase-mix its
way to equilibrium.

\subsection{Equilibrium Initial Conditions: The Plummer Sphere}

Our first full-scale test-bed calculation is of an equilibrium star cluster,
specifically a Plummer sphere.  The distribution function for this model
is given by 
\be{plummerdf}
f(E)=\left \{\begin{array}{ll}
	A|E|^{7/2} & \mbox{if $E<0$} \\
	0 & \mbox{otherwise}
	\end{array}
	\right. 
\ee
where 
\be{eperm}
E=\frac{1}{2}\left (v_r^2 +\frac{j^2}{r^2}\right )
-\frac{GM}{L}\left (1+\frac{r^2}{L^2}\right )^{-1/2}
\ee
is the energy per unit mass,
$A=\frac{256}{35\pi^4}L^2/G^5 M^5$, and $M$ is the total mass of the 
system.

Phase space is divided into 
$N_j$ ($r,v_r$) planes of constant $j$ with each plane having
$N^2$ grid points.  For the simulations that
follow we set $N=1024$ and $N_j=25$, with no $j=0$ plane.
(The \dist function is sharpley peaked in $j^2$ at the origin, and it 
is difficult to consistently integrate ${\cal F}$ with a small value for
$N_j$.
The lack of a $j=0$ plane in numerically consistent with a large value of $N_j$,
where the $j=0$ plane would have a very small weight.)
The planes are spaced equally in $j$
although in general we can select planes at arbitrary values of $j$.
We sample the set of $N_j\times N^2$,
$\{r,v_r\}_i$ phase space points using the usual N-Body technique (see e.g.
H\'enon 1966). Figure \ref{fig:sam1} illustrates this sampling method.
Here the set of sampled phase space points are shown for the
$j=0.5$ and $j=1.0$ planes. 
With these points in hand one can algorithmically compute the
coherent state summation to produce the set of $N_j$ initial wave functions.
Figure \ref{fig:plum1} 
shows the results of this construction. Here a typical phase space plane
is shown, along with the generating wave function and the mass and velocity
distributions for that plane. One should note that with the above choice
of $\eta$, the $r$ and $v_r$ resolutions are very similar. If $\eta$ were
increased, corresponding to a larger smoothing window, the mass \dist
would become smoother while the velocity \dist would become more jagged.
In terms of the \sch particles of Figure \ref{fig:particles1}, the unit
phase space particle would transform from a circle to a ellipse,
becoming shrunken in the $v_r$ direction and 
elongated in the $r$ direction. All of these effects would be transposed for
$r$ and $v_r$ if $\eta$ were, instead, to be decreased.

The system is evolved by solving the coupled Schr\"odinger-Poisson system using the
Visscher algorithm along with a potential solving routine based on
Eastwood \& Brownrigg (1979).
Care must be taken to obtain the correct boundary condition at infinity
when solving the Poisson equation 
on a finite grid. The routine that
was implemented makes use of `padding' to eliminate the effects of the grid
boundary (at the expense of doubling the number of grid points).
Figure \ref{fig:dist1} shows the mass and velocity distributions of
the system in its initial state and 
after being evolved through $10$ dynamical times.  Figure
\ref{fig:plum2} shows the $j$ plane of figure \ref{fig:plum1} after
$10$ dynamical times. From these plots alone the system does appear to
be stable. A quantitative measure of this is the virialization of the
system: for a perfect \PS $2T+W=0$.
Where $T$ is the total kinetic energy:
\beg
T=\frac{1}{2}\langle v_r^2\rangle+\sum_{j^2}j^2\int_{0}^{\infty}\left(
\frac{1}{r^2}\right)
|\psi_j|^2~dr~,
\ee
and $W$ is the potential energy of the system.
The ratio $2T/|W|$ was calculated
during the evolution of the system over the course of $50$ dynamical
times.  Figure \ref{fig:bounce1} shows the evolution of $2T/|W|$ for
the system, with a similar curve obtained using a standard treecode
(Barnes \& Hut 1986) for reference.  As can be seen from the plot, the
SM \PS appears to be on par with the particle simulation.

\subsection{Non-equilibrium Initial Conditions}

As we saw in the last section, the \PS can be modeled quite well with 
only 25 $j$ planes. But how well can this method model a 
perturbed system? One can imagine a situation where the numerical 
representation of a system is sufficient near equilibrium but is
too coarse to accurately model an evolving system. Therefore,
in order to test the true dynamics of the spherical
SM we performed a destabilized
Plummer sphere simulation (Rasio, Shapiro, \& Teukolsky (1989) use
this system to test their phase space code). 
The system was destabilized by reducing the $v_r$ 
components of all the sampled \sch particles by a factor of 
$1/\sqrt{8}$, giving $2T/|W|=17/24\simeq 0.708$. 
This was also done for the BH simulation, again with $N=1024$ 
particles. Figure \ref{fig:bounce2}
shows the evolution of $2T/|W|$ for both
of these simulations. As can be seen, both undergo the expected bounce and
then begin to virialize and settle into a new equilibrium.  
Figure \ref{fig:phasemix2}
follows one of the $j$ planes through some of this process to illustrate 
the phase mixing that occurs.

\section{Axion Dynamics and the Formation of Soliton Stars}

The SM is applicable to any collisionless system regardless of
what form the constituent particles take.  (The same holds for
N-body and phase space methods.)  This is particularly relevant for 
studies of galactic halos and large scale structure where the
dominant component of the mass density, the so-called dark matter,
is in some unknown form.  Dark matter candidates range from
astrophysical compact objects such as black holes and brown dwarfs to elementary
particles such as neutralinos.  Still the dynamics of the system
is independent of nature's choice.

There is at least one exception to the above argument.  If dark matter
is in the form of a very light scalar field, then
``quantum-mechanical'' effects will enter into the dynamical equations
of motion.  In this case, the Schr\"odinger-Poisson pair provides the
exact equations of motion (for non-relativistic motion)
for the system.  The mass $m$ is now a
true physical parameter determined by particle physics theory.
Of course so long as the Compton wavelength for the field is 
much less than the scales of interest in the problem, the field will
behave like collisionless matter.  

Recently there has been considerable interest in soliton stars
(for a review, see, e.g., Jetzer 1992).  These
objects are essentially self-gravitating compact objects made up of
bosonic fields.  The interest in soliton stars has been generated
largely by the conjecture that dark matter may be bosonic.  Indeed one
of the most popular dark matter candidates is the axion, a 
pseudo-Nambu-Goldstone boson that arises in the PQ (Peccei \& Quinn 1977)
solution to the strong CP problem (Weinberg 1978; Wilczek 1978).
Axions that are viable (there are constraints on the
axion mass from cosmology, stellar evolution, and supernova studies
(see, e.g., Kolb \& Turner 1990))
have a mass $\sim 10^{-5}\,{\rm eV}$ and therefore a de Broglie
wavelength in a galactic halo of $\sim 10\,{\rm m}$.  All this
suggests that axions behave like any other dark matter candidate,
i.e., like collisionless matter.  However Seidel and Suen (1994)
suggest that axions can form soliton stars.  The question of 
whether this is indeed the case, or whether axions, like any other
form of collisionless matter, form virialized clumps of matter,
as in the axion miniclusters of Hogan and Rees (1988), is of fundamental
importance in understanding whether or not axions are a viable
dark matter candidate.  Axions can annihilate into photons
($AA\to\gamma\gamma$) provided the density is high enough.  In the
axion miniclusters of Hogan and Rees (1988) the density is $10^7
{\rm g/cm^3}$ and annihilation is unimportant.  Alternatively
the density in axion stars can be as high as $10^{24}\,{\rm g/cm^3}$.
At these densities, the annihilation is very efficient making the
configuration unstable.  

The question of course is whether axion stars ever form in the early
Universe.  A diffuse axion cloud must loose both angular momentum and
energy if it is to form an axion star.  Seidel and Suen (1994) find
that a collapsing cloud can loose energy by radiating scalar material,
a process they call gravitational cooling.  However they only briefly
mention the issue of angular momentum and carry out simulations of
purely radial collapse.  In these simulations the field collapses
quickly and an excited soliton star forms at the center and settles
quickly by radiating scalar matter.

In what follows we argue that any reasonable amount of angular
momentum will prevent the object from reaching the stage where
gravitational cooling can take place.  
Consider an initial cloud of mass $M$, radius $R_i$, binding energy
$E_i\sim GM^2/R$, and angular momentum $J$.  A useful quantity 
is the dimensionless parameter
\be{lamdefined}
\lambda\equiv\frac{J|E|^{1/2}}{GM^{5/2}}
\ee
which is roughly the square root of the ratio between the rotational
energy and the binding energy.  As an object collapses, $\lambda\propto
R^{-1/2}$ provided there is no loss of angular momentum.  
We expect that this will indeed be the case at least in the initial 
collapse of the axion cloud.  $\lambda_f\la 1$ in order for
the final object to be gravitationally bound.  We therefore require
$\lambda_i \left (R_i/R_f\right )^{1/2}\la 1$ where $R_f\simeq
\hbar^2/\mu^2 GM$ is the radius of the final compact object.
The bound is then
\be{boundonlambda}
\lambda_i\la 10^{-12}
\left (\frac{10^{-5}{\rm eV}}{m}\right )
\left (\frac{M_\odot}{M}\right )^{2/3}
\left (\frac{\rho}{\rho_{\rm halo}}\right )^{1/6}
\ee
where $\rho_{\rm halo}=0.008 M_\odot/pc^3$ is characteristic
of the density of dark matter in the Galaxy.  This is an 
incredibly small value for $\lambda$.  It is widely 
believed that galaxies acquire angular momentum during formation
from tidal interactions with nearby mass distributions.  
Typical values from both analytic and numerical work suggest
that $\lambda\simeq 0.01-0.1$ for galaxy-sized objects (see, e.g., Peebles 1969;
Efstathiou \& Jones 1979).  The 
conclusion is that axionic matter, if indeed it does make up
the dark halo, will have too much angular momentum to ever 
reach the densities required for gravitational cooling and
soliton star formation.

\section{Summary and Conclusion}

This paper describes test-bed simulations of collisionless,
self-gravitating matter using the SM.  Included are several
improvements over the earlier work by Widrow and Kaiser (1993).  In
particular we describe how to set up ``hot'' (e.g., virialized)
systems.  We also develop the techniques required to handle
spherically symmetric systems with angular momentum which may be
useful for studying violent relaxation and equilibrium star clusters.
Of course the method is not restricted to problems with high degrees
of symmetry.  Our focus on spherically symmetric systems is
mainly for illustrative purposes and, in hindsight, may not have been
the best choice because of the special features of spherical
coordinates.  Widrow and Kaiser (1993) ran 2D cosmological simulations
with results that were in excellent agreement with those from N-body
simulations.

As discussed above, the SM is as efficient as a particle-mesh code
in following the evolution of a system in phase space.  One obvious
improvement would be to use an adaptive grid and adaptive timestep
to do the simulation.  This could in principle make the method 
competitive with the widely used treecode of Barnes and Hut (1986).

The SM can be adapted to a wide variety of problems that have, as
their central equation, Vlasov.  For example Widrow (1996) has shown
that the SM (actually the Klein-Gordon equation) can be used to
simulate collisionless systems in general relativity.  The method can
also be applied to problems in electrodynamics that require information
about the phase space structure of particles thereby providing an
alternative to particle-in-cell codes and Eulerian Vlasov codes
(see, e.g., Manfredi et al. 1995).

N-body techniques have been used in virtually all areas in
astrophysics that require numerical simulations of collisionless
matter.  However despite years of research, there are still questions
that arise over the validity and applicability of these simulations.
Most of the questions center around smoothing, the technique used to
construct a smooth density field from a system of discreet particles
(Earn \& Sellwood 1995).  The SM provides an alternative where the
``super-particles'' of the N-body experiment are replaced by a
continuous field.  While the field may look like the superposition of
particle-like wavepackets (indeed, this is how we set up our initial
conditions) these wavepackets spread out in phase space as the system
evolves.  We therefore have a type of dynamical smoothing.  Whether
this dynamical smoothing has any real advantages over N-body methods
remains to be seen.

\bigskip

{\em Acknowledgments} We would like to thank Martin Duncan and Jonathan Dursi for
helpful discussions.  Much of this work appears in a thesis
by one of us (GD) submitted in partial fulfillment of a Master's degree
at Queen's University.  This work was supported in part by a grant from
the Natural Sciences and Engineering Research Council of Canada.

\bigskip

\newpage

\noindent {\Large\bf References}

\medskip

\noindent Barnes, J. \& Hut, P. 1986, Nature {\bf 324}, 446

\medskip

\noindent Binney, J. \& Tremaine, S. 1987, Galactic Dynamics
(Princeton: Princeton University Press).

\medskip

\noindent Earn, D. J. D. \& Sellwood, J. A. 1995, ApJ {\bf 451}, 531

\medskip

\noindent Eastwood, J.W. \& Brownrigg 1979, J. Comp. Phys. {\bf 32}, 24

\medskip

\noindent Efstathiou, G. \& Jones, B. J. T. 1979, MNRAS {\bf 186}, 133

\medskip

\noindent H\'enon, M. 1973, A\&A {\bf 24}, 229

\medskip

\noindent Hockney, R. W. \& Eastwood, J. W. 1988 {\it Computer Simulation
Using Particles} (Bristol England: Adam Hilger)

\medskip

\noindent Hogan, C. J. \& Rees, M. J. 1988, Phys. Lett. {\bf B205}, 228

\medskip

\noindent Husimi, K. 1940, Proc. Phys. Math. Soc. Jpn. {\bf 22}, 264

\medskip

\noindent Jetzer, P. 1992, Phys. Rep. {\bf 220}, 163

\medskip

\noindent Kolb, E. W. \& Turner, M. S. 1990 {\it The Early Universe}
(Redwood City, California: Addison-Wesley Publishing Company)

\medskip

\noindent Manfredi, G. et al. 1995, J. Comp. Phys.,
121, 298

\medskip

\noindent Peccei, R. D. \& Quinn, H. R. 1977, Phys. Rev. Lett. {\bf 38}, 1440

\medskip

\noindent Peebles, P. J. E. 1969, ApJ {\bf 155}, 393

\medskip

\noindent Press, W. H., Flannery, B. P., Teukolsky, S. A., \& Vetterling,
W. T. 1986 {\it Numerical Recipes} (Cambridge: Cambridge University
Press)

\medskip

\noindent Rasio, F. A., Shapiro, S. L., \& Teukolsky, S. A. 1989, ApJ
{\bf 344}, 146.

\medskip

\noindent Seidel, E. \& Suen, W. 1994, Phys. Rev. Lett. {\bf 272}, 2516

\medskip

\noindent Skodje, R. T., Rohrs, H. W. \& VanBuskirk, J. 1989,
Phys. Rev. {\bf A40}, 2894.

\medskip

\noindent Visscher, P. B. 1991, Computers in Physics, Nov/Dec

\medskip

\noindent Weinberg, S. 1978, Phys. Rev. Lett. {\bf 40}, 223

\medskip

\noindent Widrow, L. M. \& Kaiser, K. 1993, ApJ {\bf 416}, L71

\medskip

\noindent Widrow, L. M., astro-ph/9607124

\medskip

\noindent Wilczek, F. 1978, Phys. Rev. Lett. {\bf 40}, 279

\medskip

\newpage

\section*{Figure Captions}

{\large\bf Fig. 1:}
Test Particles: Here the \sch test particles are moving in a fixed \PS
potential with $j=0$.
The solid cures are the expected phase space trajectories for the 
respective initial conditions. 
The top four plots are taken at $T=0$ and the bottom four at $T=10.2~T_d$.
Here and below velocities are measured in
units of $L/T_d$ where $T_d\equiv\sqrt{3\pi/32G\rho_0}$ (see text).

\medskip

\noindent {\large\bf Fig. 2:}
Test Particles: Here the \sch test particles are moving in a fixed \PS
potential with $j=0.5$.
The top four plots are taken at $T=0$ and the bottom four at $T=10.2~T_d$.
The solid cures are the expected phase space trajectories for the 
respective initial conditions.

\medskip

\noindent {\large\bf Fig. 3:}
Cold Distribution: The initial phase space \dist corresponds 
to the $j=0.5$ plane
of a \PS with the velocity \dist suppressed. The system is out of
equilibrium and rapidly collapses and begins to virialize.

\medskip

\noindent {\large\bf Fig. 4:}
Sampling the Plummer Sphere:
This figure shows the $j=0.5$ and $j=1.0$ planes of the \PS, each
with its `N-Body' sampled particles. 
This sampling is the last step prior to the
construction of the initial wave functions $\psi_j$.

\medskip

\noindent {\large\bf Fig. 5:}
The Equilibrium Plummer Sphere:
This figure shows the initial wave function of a \PS $j$ plane, along with
the associated phase space \dist, ${\cal F}_j(r,v_r)$.
The mass and velocity distributions were
calculated directly from ${\cal F}_j$. The solid curves are the model 
distributions.

\medskip

\noindent {\large\bf Fig. 6:}
The Equilibrium Plummer Sphere:
This figure shows the initial mass and velocity distributions of the 
constructed \PS, along with the same distributions after the system 
was evolved through ten dynamical times.
The dashed curves are the model distributions.

\medskip

\noindent {\large\bf Fig. 7:}
The Equilibrium Plummer Sphere:
This is the $j$ plane of figure \ref{fig:plum1}
after the system was evolved through ten dynamical times.

\medskip

\noindent {\large\bf Fig. 8:}
The equilibrium Plummer Sphere: 
This figure shows the evolution of $2T/|W|$ for the constructed \PS.
The two curves are the results of
the SM (line with points) and the BH tree code (with 1024 particles). 

\medskip

\noindent {\large\bf Fig. 9:}
The Destabilized Plummer Sphere:
This figure shows the evolution of $2T/|W|$ for the destabilized \PS.
The two curves are the results of
the SM (line with points) and the BH tree code (with 1024 particles). 

\medskip

\begin{figure}
\psfig{file=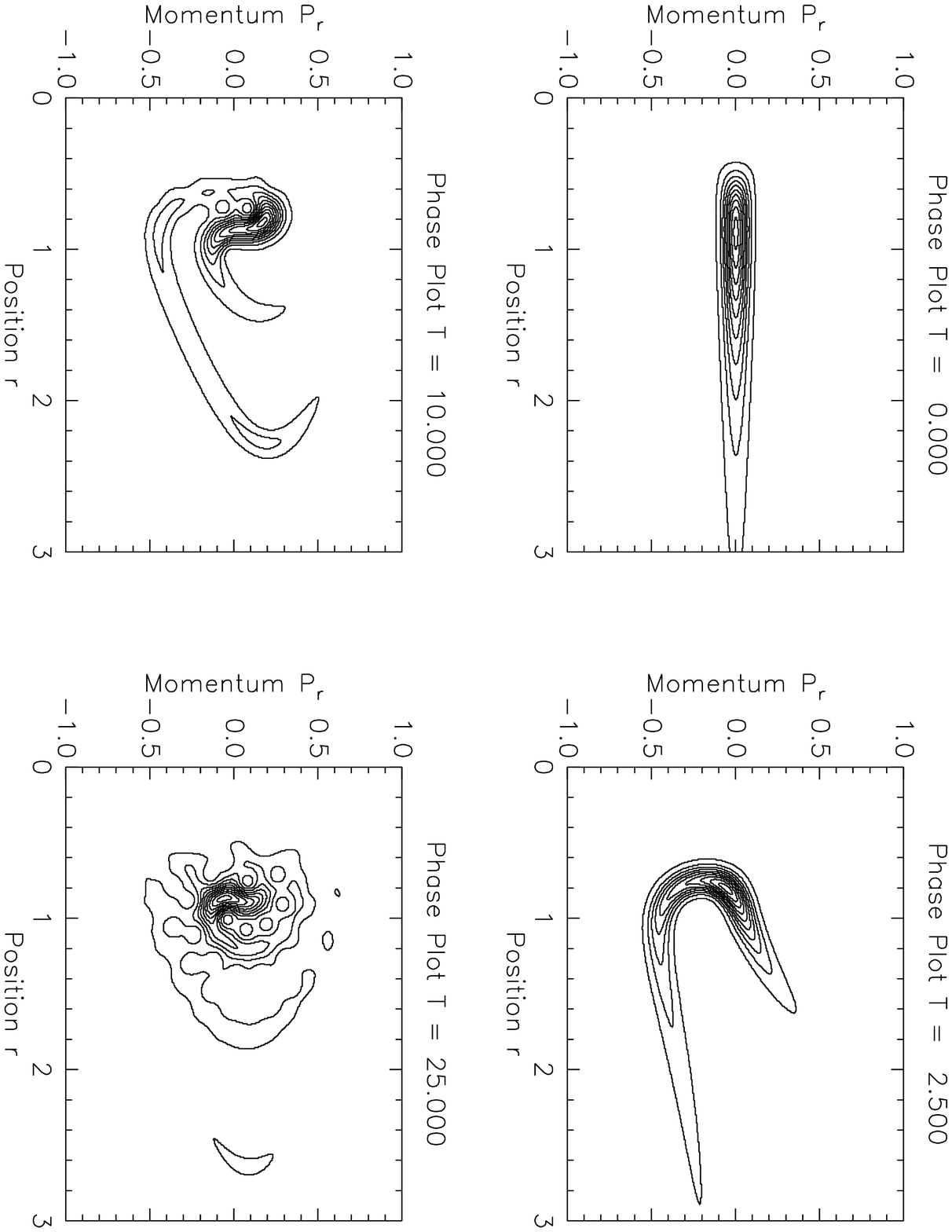,width=5.25in,angle=0}
\caption{}
\label{fig:phasemix1}
\end{figure}

\begin{figure}
\psfig{file=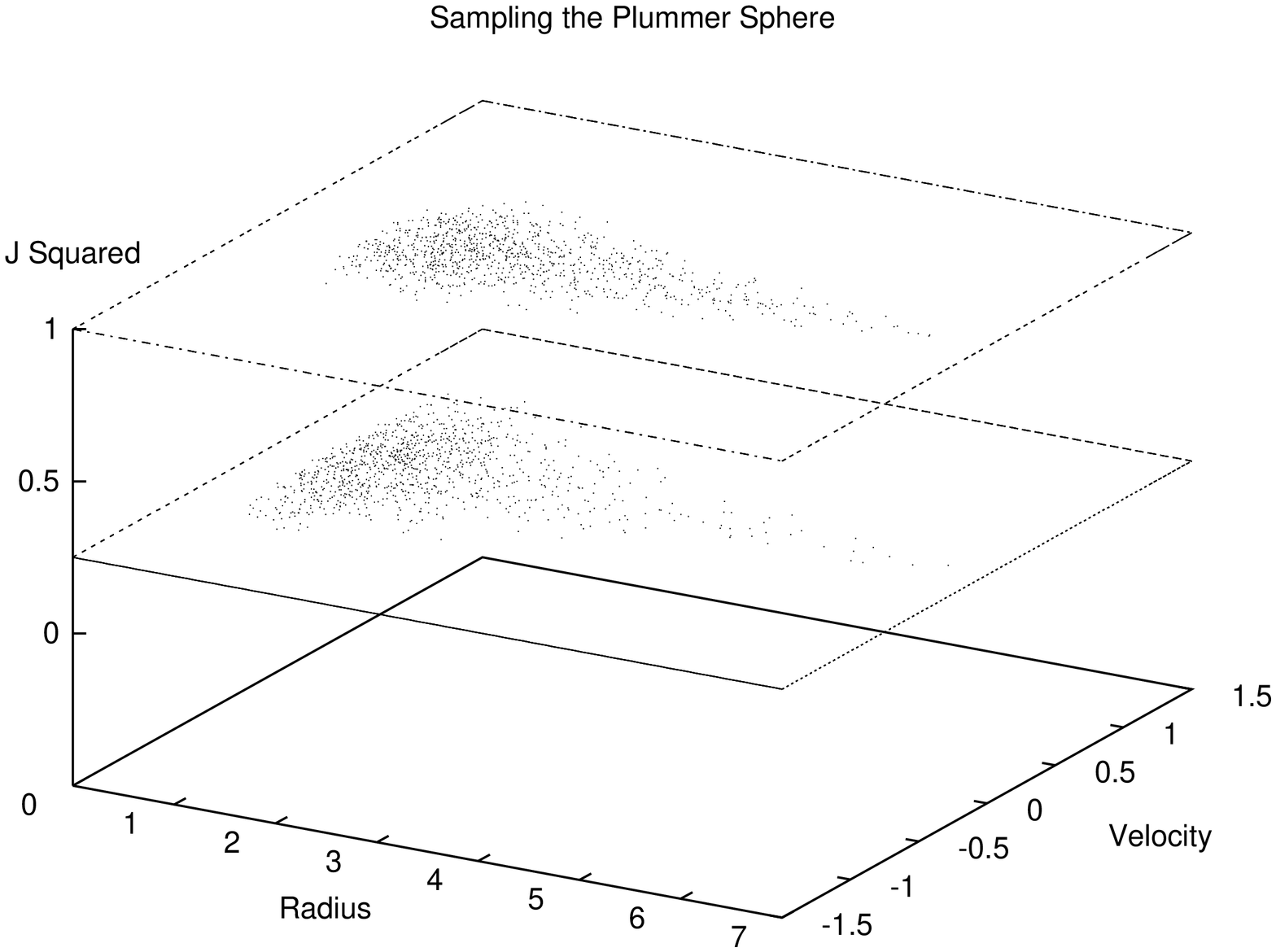,width=5in,angle=270}
\caption{}
\label{fig:sam1}
\end{figure}

\begin{figure}
\psfig{file=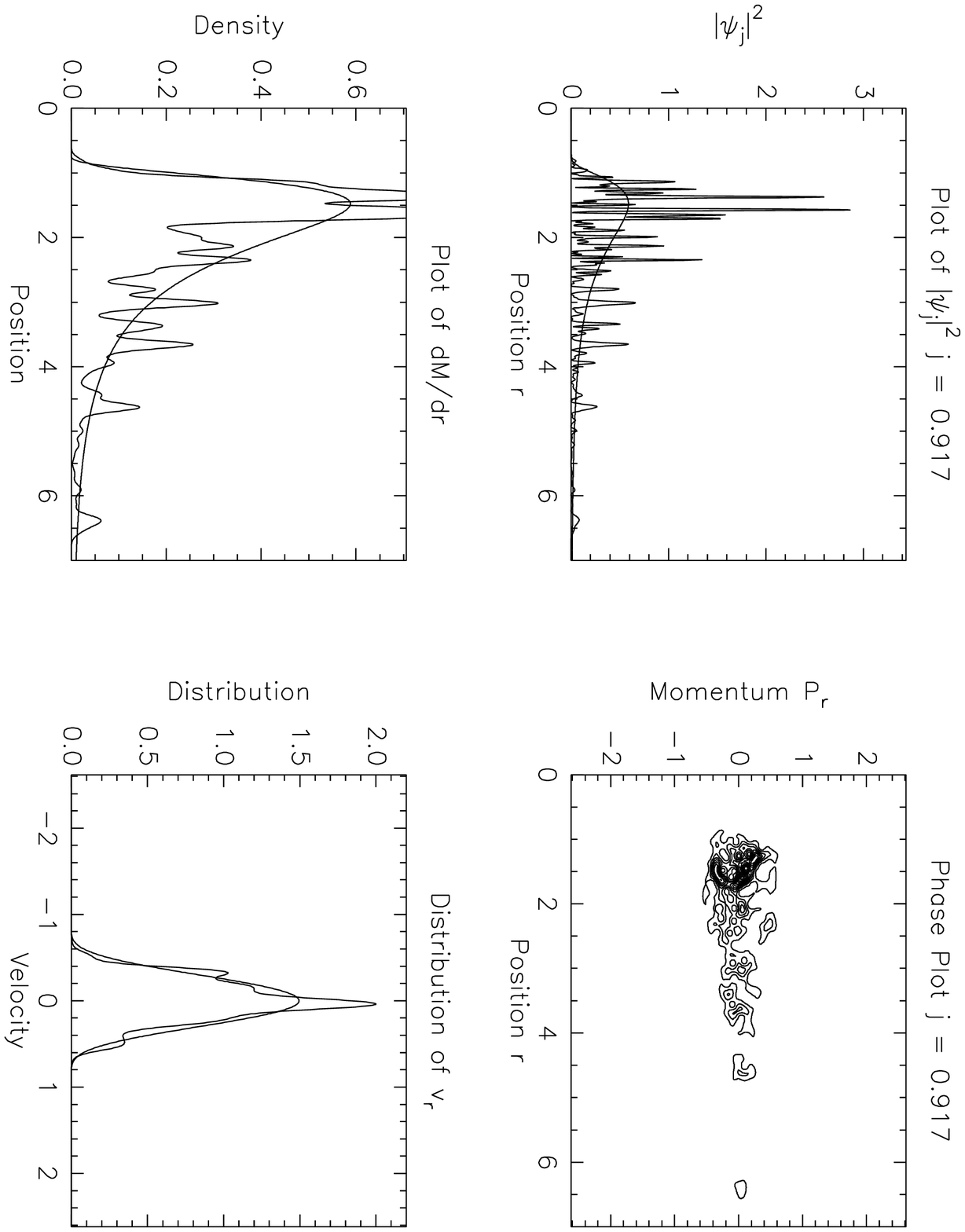,width=5in,angle=0}
\caption{}
\label{fig:plum1}
\end{figure}

\begin{figure}
\psfig{file=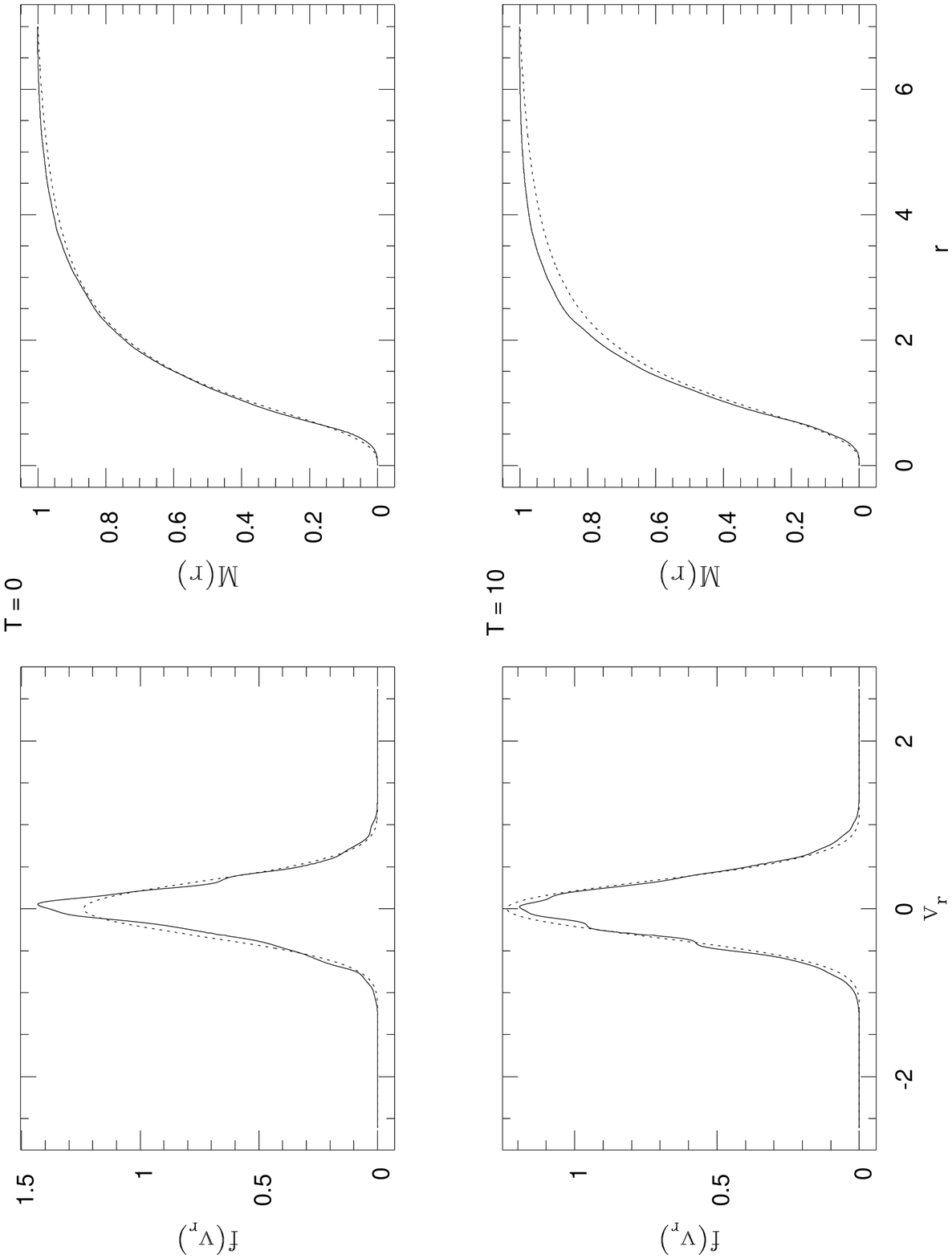,width=5in,angle=0}
\caption{}
\label{fig:dist1}
\end{figure}

\begin{figure}
\psfig{file=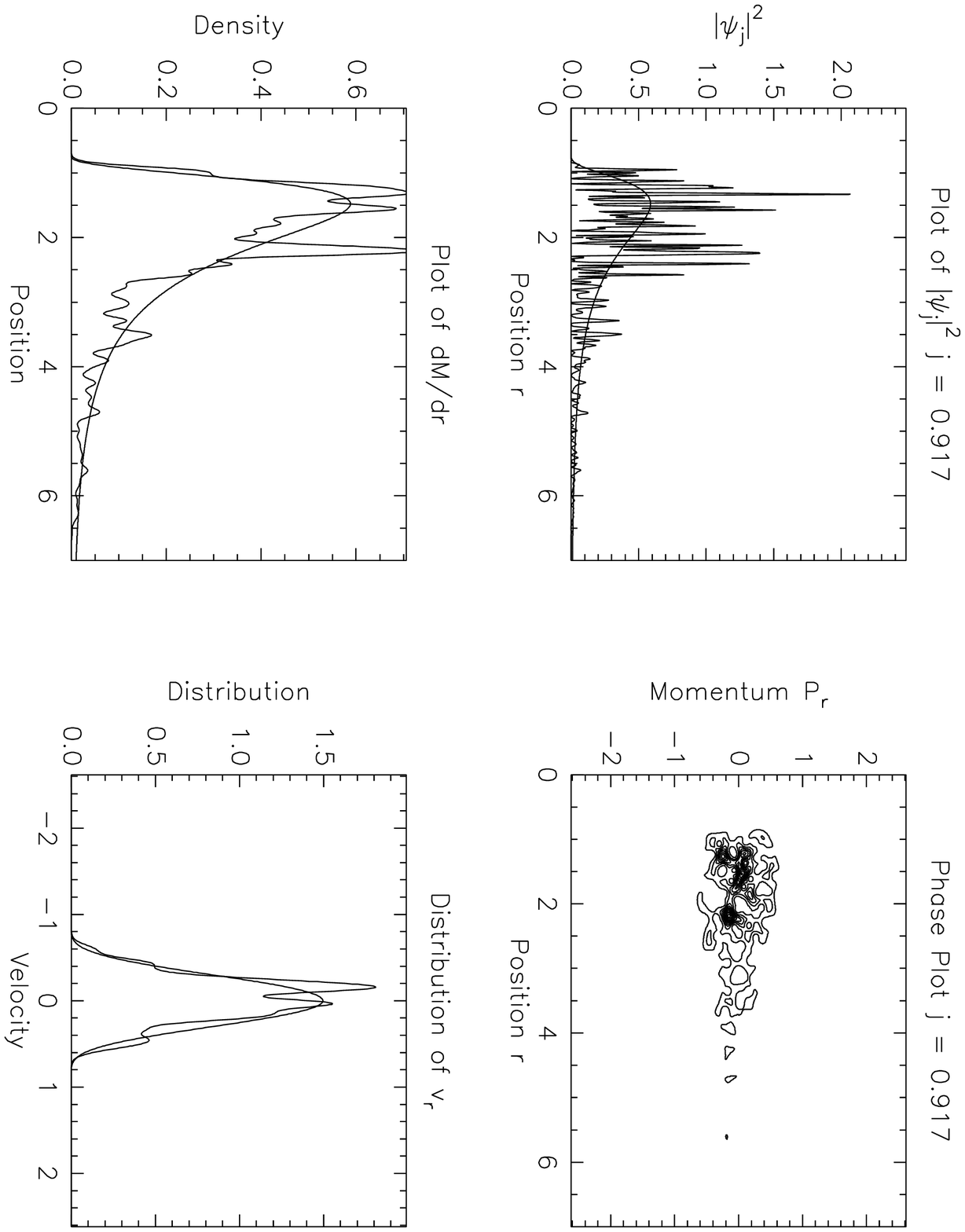,width=5in,angle=0}
\caption{}
\label{fig:plum2}
\end{figure}

\begin{figure}
\psfig{file=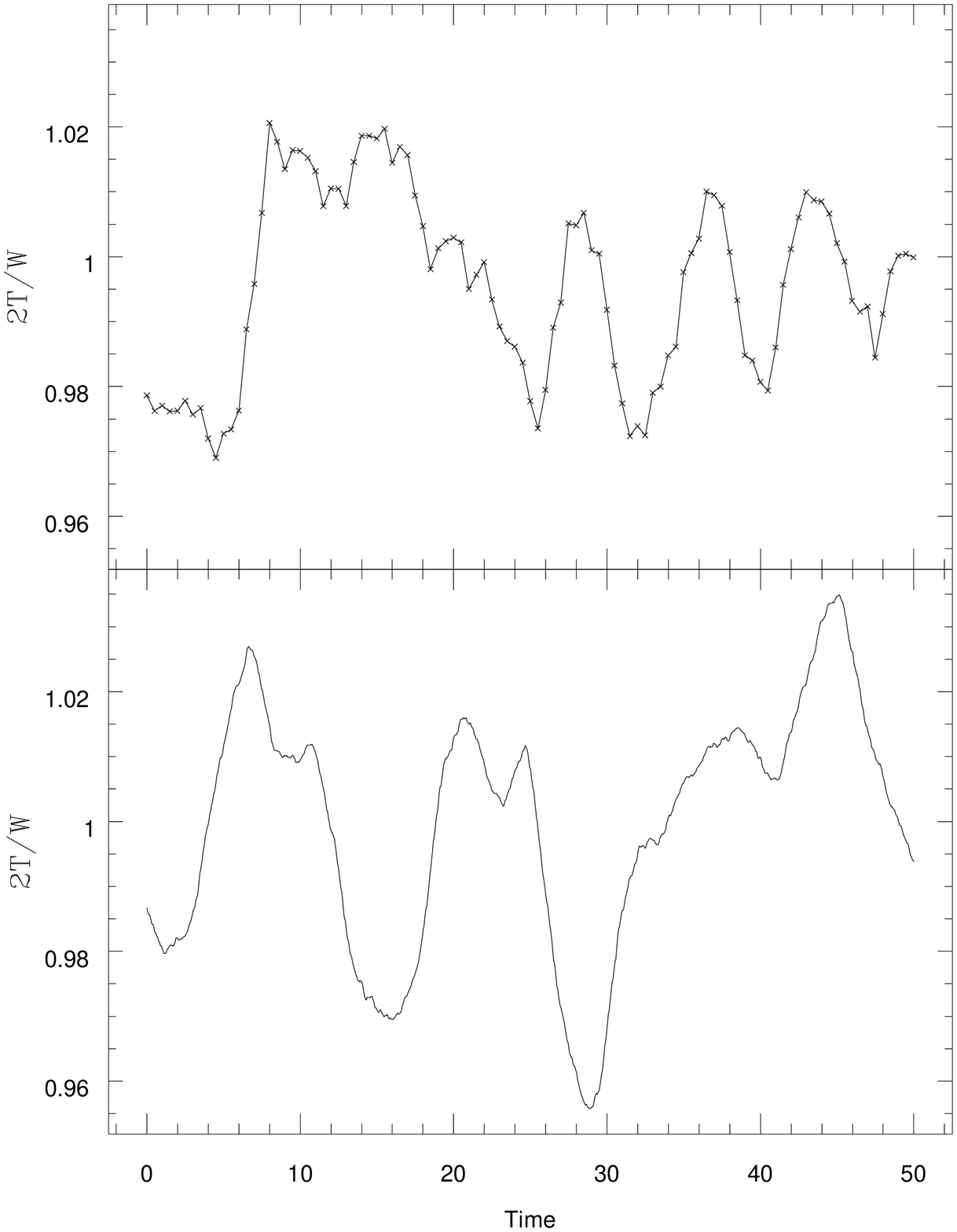,width=5in,angle=0}
\caption{}
\label{fig:bounce1}
\end{figure}

\begin{figure}
\psfig{file=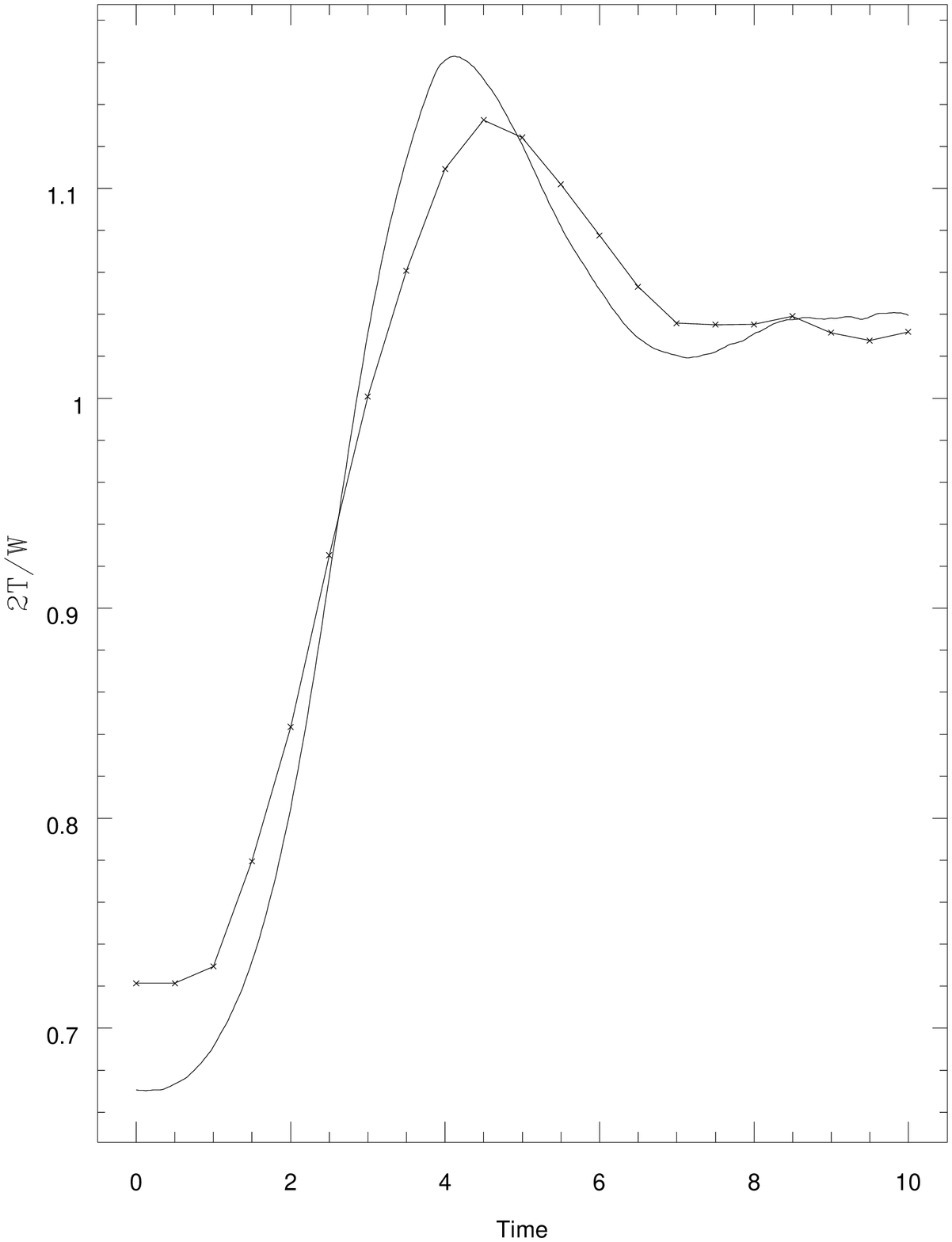,width=5in,angle=0}
\caption{}
\label{fig:bounce2}
\end{figure}

\begin{figure}
\psfig{file=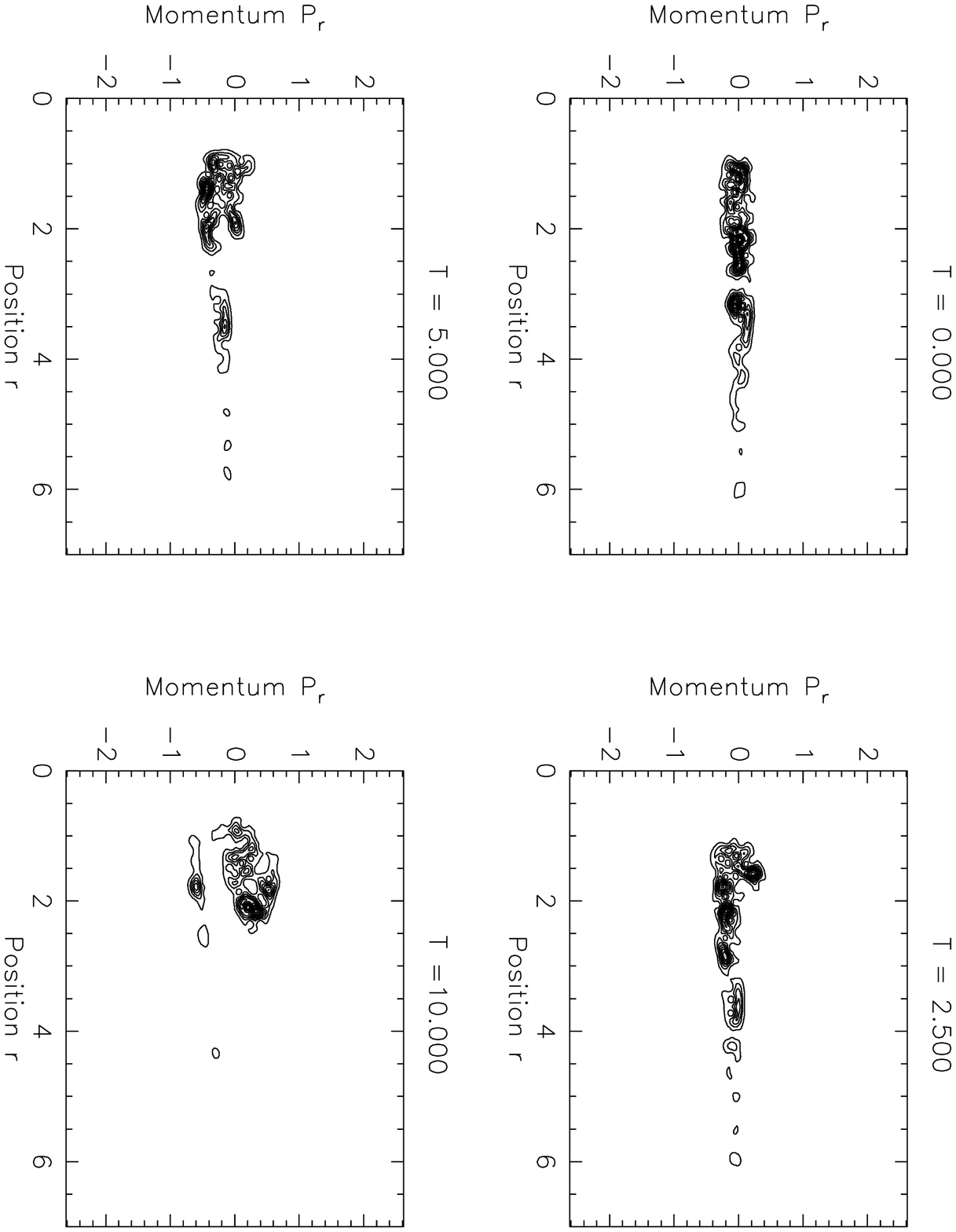,width=5.25in,angle=0}
\caption{}
\label{fig:phasemix2}
\end{figure}

\noindent {\large\bf Fig. 10:}
The Destabilized Plummer Sphere:
This figure shows the evolution of $j$ plane of the destabilized \PS.
One can see that the system is phase mixing in a similar fashion to the
cold system of figure \ref{fig:phasemix1}.

\newpage

\newpage

\newpage

\newpage

\newpage

\newpage

\newpage

\newpage

\newpage

\newpage

\end{document}